\documentstyle[prd,aps,preprint,tighten,epsfig]{revtex}

\begin{document}

\draft

\title{A Shift from Democratic to Tri-bimaximal Neutrino Mixing \\
with Relatively Large $\theta^{}_{13}$}
\author{{\bf Zhi-zhong Xing}
\thanks{E-mail: xingzz@ihep.ac.cn}}
\address{Institute of High Energy Physics, Chinese Academy of Sciences,
Beijing 100049, China}

\maketitle

\begin{abstract}
Recent neutrino oscillation data hint that the smallest neutrino
mixing angle $\theta^{}_{13}$ is possible to lie in the range
$5^\circ \lesssim \theta^{}_{13} \lesssim 12^\circ$. We show that
reasonable perturbations to the democratic mixing pattern,
which is geometrically related to the tri-bimaximal mixing pattern through
an equal shift $\theta^{}_* \simeq 9.7^\circ$ of two large mixing
angles, can naturally produce a nearly tri-bimaximal neutrino
mixing matrix $V$ with sufficiently large $\theta^{}_{13}$. Two
especially simple but viable scenarios of $V$ are proposed and their
phenomenological consequences are discussed.
\end{abstract}


\newpage


\framebox{\large\bf 1} ~
Recent solar, atmospheric, reactor and accelerator neutrino
oscillation experiments have provided us with very convincing evidence
that neutrinos are massive and lepton flavors are mixed \cite{PDG10}.
The mixing of lepton flavors is effectively described by
a $3\times 3$ unitary matrix $V$, whose nine elements can be
parametrized in terms of three rotation angles and three
CP-violating phases. Defining three rotation matrices in the (1,2),
(1,3) and (2,3) planes as
\begin{eqnarray}
R^{}_{12}(\theta^{}_{12}) & = & \left(
\matrix{c^{}_{12} & s^{}_{12} & 0 \cr -s^{}_{12} &
c^{}_{12} & 0 \cr 0 & 0 & 1 \cr} \right) \; , \nonumber \\
R^{}_{13}(\theta^{}_{13}) & = & \left(
\matrix{c^{}_{13} & 0 & s^{}_{13} \cr 0 & 1 & 0 \cr
-s^{}_{13} & 0 & c^{}_{13} \cr} \right) \; , \nonumber \\
R^{}_{23}(\theta^{}_{23}) & = & \left( \matrix{1 & 0
& 0 \cr 0 & c^{}_{23} & s^{}_{23} \cr 0 & -s^{}_{23} &
c^{}_{23} \cr} \right) \; ,
\end{eqnarray}
where $c^{}_{ij} \equiv \cos\theta^{}_{ij}$ and $s^{}_{ij}
\equiv \sin\theta^{}_{ij}$ (for $1\leq i<j
\leq 3$), one may parametrize $V$ in nine topologically different
ways \cite{FX01}. The so-called standard parametrization takes the form
\begin{eqnarray}
V & = & R^{}_{23}(\theta^{}_{23}) \otimes P^{}_\delta \otimes
R^{}_{13}(\theta^{}_{13}) \otimes P^\dagger_\delta
\otimes R^{}_{12}(\theta^{}_{12}) \otimes P^{}_\nu \nonumber \\
& = & \left( \matrix{ c^{}_{12}
c^{}_{13} & s^{}_{12} c^{}_{13} & s^{}_{13} e^{-i\delta} \cr
-s^{}_{12} c^{}_{23} - c^{}_{12} s^{}_{13} s^{}_{23} e^{i\delta} &
c^{}_{12} c^{}_{23} - s^{}_{12} s^{}_{13} s^{}_{23} e^{i\delta} &
c^{}_{13} s^{}_{23} \cr s^{}_{12} s^{}_{23} - c^{}_{12} s^{}_{13}
c^{}_{23} e^{i\delta} & -c^{}_{12} s^{}_{23} - s^{}_{12} s^{}_{13}
c^{}_{23} e^{i\delta} & c^{}_{13} c^{}_{23} \cr} \right) P^{}_\nu \;
,
\end{eqnarray}
in which $P^{}_\delta = {\rm Diag}\{1, 1, e^{i\delta}\}$ and
$P^{}_\nu ={\rm Diag}\{e^{i\rho}, e^{i\sigma}, 1\}$ are two diagonal
phase matrices containing three CP-violating phases. A
recent global analysis of current neutrino oscillation data yields
$\theta^{}_{12} = 34.5^\circ \pm 1.0^\circ$,
$\theta^{}_{13} = 5.1^{+3.0^\circ}_{-3.3^\circ}$ and
$\theta{}_{23} = 42.8^{+4.7^\circ}_{-2.9^\circ}$
at the $1\sigma$ level \cite{GG}, but three phases of $V$ remain
entirely unconstrained. The ongoing and forthcoming neutrino
oscillation experiments will measure $\theta^{}_{13}$ and $\delta$.
On the other hand, the neutrinoless double-beta decay experiments
will help to probe or constrain $\rho$ and $\sigma$.

The smallness of $\theta^{}_{13}$ and the largeness of
$\theta^{}_{12}$ and $\theta^{}_{23}$ have motivated some
speculations about a constant neutrino mixing matrix with
$\theta^{}_{13} =0^\circ$, such as the ``democratic"
pattern
\begin{equation}
U^{}_0 = \left( \matrix{ \sqrt{\frac{1}{2}} & \sqrt{\frac{1}{2}}
& 0 \cr -\sqrt{\frac{1}{6}} & \sqrt{\frac{1}{6}} &
\sqrt{\frac{2}{3}} \cr \sqrt{\frac{1}{3}} & -\sqrt{\frac{1}{3}} &
\sqrt{\frac{1}{3}} \cr} \right) \;
\end{equation}
with $\theta^{(0)}_{12} = 45^\circ$, $\theta^{(0)}_{13} = 0^\circ$ and
$\theta^{(0)}_{23} =\arctan(\sqrt{2}) \simeq 54.7^\circ$ \cite{FX96} or
the ``tri-bimaximal" pattern
\begin{equation}
V^{}_0 = \left( \matrix{ \sqrt{\frac{2}{3}} & \sqrt{\frac{1}{3}}
& 0 \cr -\sqrt{\frac{1}{6}} & \sqrt{\frac{1}{3}} &
\sqrt{\frac{1}{2}} \cr \sqrt{\frac{1}{6}} & -\sqrt{\frac{1}{3}} &
\sqrt{\frac{1}{2}} \cr} \right) \;
\end{equation}
with $\vartheta^{(0)}_{12} = \arctan(1/\sqrt{2}) \simeq 35.3^\circ$, $\vartheta^{(0)}_{13} = 0^\circ$ and
$\vartheta^{(0)}_{23} = 45^\circ$ \cite{TB}.
Either of them can be obtained in the limit of a certain flavor
symmetry (e.g., the discrete S(3) flavor symmetry for $U^{}_0$ or
the discrete $A^{}_4$ symmetry for $V^{}_0$ \cite{AF00}), and the
latter has to be broken in order to generate nonzero
$\theta^{}_{13}$ and CP violation. Since $V^{}_0$ is much
closer to the best-fit values of current data on three
neutrino mixing angles, it has recently attracted much more interest.

Note that the entries of $V^{}_0$ are actually the same as those of
$U^{}_0$, although their positions are essentially different. Hence
it is interesting to explore not only an intrinsic relationship
between $U^{}_0$ and $V^{}_0$ but also how to link them to the realistic neutrino mixing matrix $V$ via reasonable perturbations. Note also that
there are some preliminary hints that the smallest neutrino mixing
angle $\theta^{}_{13}$ might not be very small. For example,
$\theta^{}_{13} \simeq 5.1^{+3.0^\circ}_{-3.3^\circ}$ ($1\sigma$)
by Gonzalez-Garcia {\it et al} \cite{GG},
$\theta^{}_{13} \simeq 7.3^{+2.0^\circ}_{-2.9^\circ}$ (1$\sigma$)
by Fogli {\it et al} \cite{Fogli},
and $\theta^{}_{13} \simeq 8.1^{+2.8^\circ}_{-4.5^\circ}$ as the best-fit
value by the KamLAND Collaboration \cite{KM}.
Although the statistical significance of these results remains quite low,
they {\it do} imply that $\theta^{}_{13}$ is possible to lie in the range
$5^\circ \lesssim \theta^{}_{13} \lesssim 12^\circ$. On the theoretical
side, it is certainly likely that $\theta^{}_{13}$ may take a value
in the above range \cite{King}. So it makes sense to discuss how to
confront a constant neutrino mixing pattern with a relatively large value of $\theta^{}_{13}$. On the other hand, a definite determination of
$\theta^{}_{13}$ may serve as a crucial turning-point of experimental
neutrino physics to the era of precision measurements, in which the
detection of leptonic CP violation and the search for new physics will
become feasible \cite{Xing2008}.

Let us pose two immediate and interesting questions: (1) what is the
geometric relation between the democratic and tri-bimaximal neutrino
mixing patterns? (2) which of them is more natural to receive
relatively significant perturbations in order to accommodate
relatively large $\theta^{}_{13}$?
In this paper we shall point out that two large mixing angles
predicted in the democratic mixing pattern $U^{}_0$ are
intrinsically related to their counterparts in the tri-bimaximal
mixing pattern $V^{}_0$ through an equal shift
\begin{equation}
\theta^{}_* \equiv \theta^{(0)}_{12} - \vartheta^{(0)}_{12} =
\theta^{(0)}_{23} - \vartheta^{(0)}_{23} \simeq 9.7^\circ \; .
\end{equation}
This geometric relation keeps unchanged if a universal perturbation
is imposed onto three mixing angles of $U^{}_0$ or $V^{}_0$.
Although it is likely to generate relatively large $\theta^{}_{13}$
by perturbing $V^{}_0$, the perturbation term has to
be adjusted in such a way that two large mixing angles
of $V^{}_0$ are slightly modified but the smallest angle of
$V^{}_0$ is significantly modified.
In contrast, it is more natural to produce
a nearly tri-bimaximal neutrino mixing matrix $V$ with
sufficiently large $\theta^{}_{13}$ by introducing comparable
perturbations to three mixing angles of $U^{}_0$. We shall propose
two especially simple but viable scenarios of $V$ --- scenario A is
based on the standard parametrization of $U^{}_0$ and scenario B
relies on a very useful parametrization proposed by Fritzsch and Xing (FX)
\cite{FX97}. Scenario A predicts $\theta^{}_{12} \simeq 35.3^\circ$,
$\theta^{}_{13} \simeq 9.7^\circ$ and $\theta^{}_{23} = 45^\circ$
together with the maximal strength of leptonic
CP violation ${\cal J}^{}_{\max} = (\sqrt{2} + 1)/ (36 \sqrt{3})
\simeq 3.9\%$; and scenario B predicts $28.3^\circ \lesssim
\theta^{}_{12} \lesssim 42.2^\circ$, $\theta^{}_{13} \simeq 6.9^\circ$
and $\theta^{}_{23} \simeq 44.6^\circ$ together with
${\cal J}^{}_{\max} = 1/36 \simeq 2.8\%$. Both scenarios are in good
agreement with current data, and they can soon be tested
in a variety of neutrino oscillation experiments.

\vspace{0.5cm}

\framebox{\large\bf 2} ~
First of all, the following relation between the democratic
mixing matrix $U^{}_0$ and the tri-bimaximal mixing matrix $V^{}_0$
comes into our notice:
\begin{eqnarray}
V^{}_0 = R^T_{23}(\theta^{}_*) \otimes U^{}_0 \otimes R^T_{12}
(\theta^{}_*) \; ,
\end{eqnarray}
where ``$T$" means a transpose, and $\theta^{}_*$ has been
defined in Eq. (5). As a matter of fact, $U^{}_0$ and $V^{}_0$
can be decomposed into
\begin{eqnarray}
U^{}_0 & = & R^{}_{23}(45^\circ + \theta^{}_*) \otimes
R^{}_{12}(45^\circ) \; ,
\nonumber \\
V^{}_0 & = & R^{}_{23}(45^\circ) \otimes
R^{}_{12}(45^\circ - \theta^{}_*)
\; .
\end{eqnarray}
Two nonzero mixing angles of $U^{}_0$ turn out to be
$\theta^{(0)}_{12} = 45^\circ$ and
$\theta^{(0)}_{23} = 45^\circ + \theta^{}_*$,
and those of $V^{}_0$ are $\vartheta^{(0)}_{12} = 45^\circ -\theta^{}_*$
and $\vartheta^{(0)}_{23} = 45^\circ$. Their geometrical relations
in the real plane are shown in FIG. 1.
So $\theta^{(0)}_{12}$ and $\theta^{(0)}_{23}$ are intrinsically related
to $\vartheta^{(0)}_{12}$ and $\vartheta^{(0)}_{23}$
via an equal shift $\theta^{}_* \simeq 9.7^\circ$. Although the size
of $\theta^{}_*$ is not small, it is smaller than the Cabibbo angle of
quark mixing (i.e., $\theta^{}_{\rm C} \simeq 13^\circ$
\cite{PDG10}). In this sense we argue that $V^{}_0$ can be regarded
as a consequence of $U^{}_0$ whose (1,2) and (2,3) mixing angles are
corrected by $\theta^{}_*$ in a destructive way.

Since $V^{}_0$ itself is very close to the best-fit result obtained
from current experimental data on three
neutrino mixing angles \cite{GG}, any possible perturbations to $V^{}_0$
must be small enough (see, e.g., Refs. \cite{Rodejohann,King2,Smirnov}).
In the standard parametrization the overall perturbation
matrix can be expressed as
\begin{eqnarray}
\Omega^{}_\varepsilon =
R^T_{23}(\varepsilon^{}_{23}) \otimes P^{}_\delta \otimes
R^{}_{13}(\varepsilon^{}_{13}) \otimes P^\dagger_\delta \otimes
R^T_{12}(\varepsilon^{}_{12}) \;
\end{eqnarray}
with $|\varepsilon^{}_{ij}| \ll 1$ (for $ij=12, 13, 23$),
and thus the overall neutrino mixing matrix is given by
\begin{eqnarray}
V & = & R^{}_{23}(45^\circ) \otimes \Omega^{}_\varepsilon \otimes R^{}_{12}
(45^\circ - \theta^{}_*) \otimes P^{}_\nu
\nonumber \\
& = & R^{}_{23}(45^\circ - \varepsilon^{}_{23}) \otimes P^{}_\delta \otimes
R^{}_{13}(\varepsilon^{}_{13}) \otimes P^\dagger_\delta \otimes
R^{}_{12}(45^\circ - \theta^{}_* - \varepsilon^{}_{12}) \otimes P^{}_\nu \;
\end{eqnarray}
with $\theta^{}_{12} = 45^\circ -\theta^{}_* -\varepsilon^{}_{12}$,
$\theta^{}_{13} = \varepsilon^{}_{13}$ and
$\theta^{}_{23} = 45^\circ -\varepsilon^{}_{23}$.
In view of $\theta^{}_{12} = 34.5^\circ \pm 1.0^\circ$,
$\theta^{}_{13} = 5.1^{+3.0^\circ}_{-3.3^\circ}$
and $\theta{}_{23} = 42.8^{+4.7^\circ}_{-2.9^\circ}$ ($1\sigma$) extracted
from a global analysis of current neutrino oscillation data \cite{GG},
one immediately obtains $\varepsilon^{}_{12} = 0.8^\circ \pm 1^\circ$,
$\varepsilon^{}_{13} = 5.1^{+3.0^\circ}_{-3.3^\circ}$
and $\varepsilon^{}_{23} = 2.2^{+4.7^\circ}_{-2.9^\circ}$. One might
argue that it would be somewhat unnatural if a perturbation to
the smallest mixing angle of $V^{}_0$ were much larger than the ones
to two large angles of $V^{}_0$. In this sense
$|\varepsilon^{}_{13}| \lesssim |\varepsilon^{}_{12}|
\lesssim |\varepsilon^{}_{23}|$ seems to be a natural choice of
three perturbation parameters, just corresponding to the fact
$\theta^{}_{13} < \theta^{}_{12} < \theta^{}_{23}$.
Then $\theta^{}_{13} = \varepsilon^{}_{13}$ is expected
to be very small, and it is most likely to lie in the range
$0^\circ \lesssim \theta^{}_{13} < 5^\circ$. In other words,
it seems rather unlikely to obtain $\theta^{}_{13} \gtrsim 5^\circ$
by introducing {\it natural} perturbations to three mixing angles of
$V^{}_0$. Note, however, that such arguments might not work when
the neutrino mixing matrix is derived from a lepton mass model.
From the point of view of model building, one is not subject to the
assumption of $|\varepsilon^{}_{13}| \lesssim |\varepsilon^{}_{12}|
\lesssim |\varepsilon^{}_{23}|$ because three mixing angles may receive
contributions from both the charged-lepton and neutrino sectors at
the tree level \cite{Rodejohann2} and they can also receive appreciable
quantum corrections at the loop level \cite{Araki}.

Different from $V^{}_0$, $U^{}_0$ is not so close to the best-fit 
neutrino mixing pattern extracted from a global analysis of current
neutrino oscillation data. Hence large perturbations to three mixing 
angles of $U^{}_0$ can naturally be allowed, in order to bring 
$U^{}_0$ to a phenomenologically favored form. In this case even
$\theta^{}_{13} \sim \theta^{}_*$ may be achieved from reasonable
perturbations to $U^{}_0$, as we shall see later on.

\vspace{0.5cm}

\framebox{\large\bf 3} ~
We proceed to discuss reasonable perturbations to the democratic
mixing pattern $U^{}_0$ so as to produce a realistic neutrino
mixing matrix $V$ with naturally large $\theta^{}_{13}$. To illustrate,
we propose two simple but viable scenarios of $V$ in two different
parametrizations of $U^{}_0$. They are based on the
standard and FX parametrizations of $U^{}_0$, leading respectively
to the predictions $\theta^{}_{13} \simeq 9.7^\circ$ and
$\theta^{}_{13} \simeq 6.9^\circ$ in the standard parametrization of $V$.

\begin{center}
{\bf Scenario A}
\end{center}

The standard parametrization of $U^{}_0$ has been given in Eq. (7).
Here we assume a universal angle
$\varepsilon^{}_{12} = \varepsilon^{}_{13} = \varepsilon^{}_{23} =
\theta^{}_*$ to perturb three mixing angles of $U^{}_0$, in order to
obtain a nearly tri-bimaximal neutrino mixing matrix $V$ with
sufficiently large $\theta^{}_{13}$. Such a perturbation term is
reasonable in the sense that its magnitude is actually smaller than
the Cabibbo angle $\theta^{}_{\rm C}$.
Let us recall that the realistic Cabibbo-Kobayashi-Maskawa quark mixing
matrix \cite{PDG10} can be regarded as a result obtained from
small perturbations to the identity matrix, in which the maximal
perturbation term is just characterized by $\sin\theta^{}_{\rm C}$.
Now a realistic neutrino mixing matrix $V$ arises from a universal
perturbation of ${\cal O}(\theta^{}_*)$ to three mixing angles of
$U^{}_0$. In this case the overall perturbation matrix reads
\begin{eqnarray}
\Omega^{}_* = R^T_{23}(\theta^{}_*) \otimes P^{}_\delta \otimes
R^{}_{13}(\theta^{}_*) \otimes P^\dagger_\delta \otimes
R^T_{12}(\theta^{}_*) \; .
\end{eqnarray}
The resultant neutrino mixing matrix is
\begin{eqnarray}
V & = & R^{}_{23}(45^\circ + \theta^{}_*) \otimes \Omega^{}_* \otimes
R^{}_{12} (45^\circ) \otimes P^{}_\nu
\nonumber \\
& = & R^{}_{23}(45^\circ) \otimes P^{}_\delta
\otimes R^{}_{13}(\theta^{}_*) \otimes P^\dagger_\delta \otimes
R^{}_{12}(45^\circ - \theta^{}_*) \otimes P^{}_\nu
\nonumber \\
& = & \left( \matrix{
\sqrt{\frac{2}{3}} \hspace{0.1cm} c^{}_* & \sqrt{\frac{1}{3}}
\hspace{0.1cm} c^{}_* & s^{}_* e^{-i\delta} \cr
-\sqrt{\frac{1}{6}} - \sqrt{\frac{1}{3}} \hspace{0.1cm} s^{}_* e^{i\delta} & \sqrt{\frac{1}{3}} - \sqrt{\frac{1}{6}} \hspace{0.1cm} s^{}_* e^{i\delta} &
\sqrt{\frac{1}{2}} \hspace{0.1cm} c^{}_* \cr
\sqrt{\frac{1}{6}} - \sqrt{\frac{1}{3}} \hspace{0.1cm} s^{}_* e^{i\delta} & -\sqrt{\frac{1}{3}} - \sqrt{\frac{1}{6}} \hspace{0.1cm} s^{}_* e^{i\delta} &
\sqrt{\frac{1}{2}} \hspace{0.1cm} c^{}_* \cr} \right) P^{}_\nu \; ,
\end{eqnarray}
where $c^{}_* \equiv \cos\theta^{}_* = (\sqrt{2} +1)/\sqrt{6}$ and
$s^{}_* \equiv \sin\theta^{}_* = (\sqrt{2} -1)/\sqrt{6}$.
This ansatz apparently predicts
$\theta^{}_{12} = 45^\circ - \theta^{}_* \simeq 35.3^\circ$,
$\theta^{}_{13} = \theta^{}_* \simeq 9.7^\circ$ and
$\theta^{}_{23} = 45^\circ$. So it is a nearly tri-bimaximal neutrino
mixing pattern with a very appreciable value of $\theta^{}_{13}$.
The Jarlskog parameter of leptonic CP violation \cite {J} is given by
\begin{eqnarray}
{\cal J} = c^{}_{12} s^{}_{12} c^2_{13} s^{}_{13} c^{}_{23} s^{}_{23} \sin\delta = \frac{\sqrt{2} + 1}{36 \sqrt{3}} \sin\delta
\lesssim 0.039 \sin\delta \;
\end{eqnarray}
in this scenario. The relatively large $\theta^{}_{13}$ and (perhaps)
$|{\cal J}|$ make scenario A easily testable in a variety of
neutrino oscillation experiments in the near future.

\begin{center}
{\bf Scenario B}
\end{center}

In the FX parametrization a generic $3\times 3$ neutrino mixing matrix
$V$ is expressed as
\begin{eqnarray}
V & = & R^{}_{12}(\theta^{}_l) \otimes R^{}_{23}(\theta) \otimes
P^\dagger_\phi \otimes R^T_{12}(\theta^{}_\nu) \otimes P^{}_\nu
\nonumber \\
& = & \left ( \matrix{ s^{}_l s^{}_{\nu} c + c^{}_l c^{}_{\nu}
e^{-i\phi} & ~ s^{}_l c^{}_{\nu} c - c^{}_l s^{}_{\nu} e^{-i\phi} ~ &
s^{}_l s \cr c^{}_l s^{}_{\nu} c - s^{}_l c^{}_{\nu} e^{-i\phi} &
c^{}_l c^{}_{\nu} c + s^{}_l s^{}_{\nu} e^{-i\phi} & c^{}_l s \cr
- s^{}_{\nu} s   & - c^{}_{\nu} s   & c \cr } \right ) P^{}_\nu \; ,
\end{eqnarray}
where $P^{}_\phi = {\rm Diag}\{e^{i\phi}, 1, 1\}$ and
$P^{}_\nu = {\rm Diag}\{e^{i\rho}, e^{i\sigma}, 1\}$ together with
$c^{}_l \equiv \cos\theta^{}_l$, $s^{~}_l \equiv
\sin\theta_l$, $c^{}_\nu \equiv \cos\theta^{}_\nu$, $s^{}_{\nu}
\equiv \sin\theta^{}_{\nu}$, $c \equiv \cos\theta$ and $s \equiv
\sin\theta$. This representation of $V$ has proved to be more
convenient and useful than the standard one in deriving the
one-loop renormalization-group
equations of three neutrino mixing angles and three CP-violating phases
\cite{Xing05} and in linking flavor mixing parameters to the ratios of
charge-lepton and neutrino masses \cite{FX06}. It coincides with
the standard parametrization in the $\theta^{}_l \to 0$ limit (up to
a rearrangement of the phase convention), in
which $\theta^{}_{12} = \theta^{}_\nu$ and $\theta^{}_{23} = \theta$
exactly hold. Hence the democratic mixing pattern $U^{}_0$ can also
be decomposed into a product of $R^{}_{23}(45^\circ + \theta^{}_*)$ and $R^{}_{12}(45^\circ)$ in the FX parametrization with
$\theta^{}_\nu = 45^\circ$ and $\theta = 45^\circ + \theta^{}_*$.
Here again we assume a universal angle $\theta^{}_*$ to perturb three
mixing angles of $U^{}_0$ in the form of
\begin{eqnarray}
~~\Omega^{}_l = R^{}_{12}(\theta^{}_*) ~~~{\rm and}~~~
\Omega^{}_\nu = R^T_{23}(\theta^{}_*) \otimes
P^\dagger_\phi \otimes R^T_{12}(\theta^{}_*) \; .
\end{eqnarray}
Then we obtain
\begin{eqnarray}
V & = & \Omega^{}_l \otimes R^{}_{23}(45^\circ + \theta^{}_*) \otimes \Omega^{}_\nu \otimes R^{}_{12}(45^\circ) \otimes P^{}_\nu
\nonumber \\
& = & R^{}_{12}(\theta^{}_*) \otimes R^{}_{23}(45^\circ) \otimes
P^\dagger_\phi  \otimes R^{}_{12}(45^\circ - \theta^{}_*) \otimes P^{}_\nu
\nonumber \\
& = & \left ( \matrix{ \sqrt{\frac{1}{6}} \hspace{0.1cm} s^{}_* +
\sqrt{\frac{2}{3}} \hspace{0.1cm} c^{}_* e^{-i\phi} & \sqrt{\frac{1}{3}} \hspace{0.1cm} s^{}_* - \sqrt{\frac{1}{3}} \hspace{0.1cm} c^{}_*
e^{-i\phi} & \sqrt{\frac{1}{2}} \hspace{0.1cm} s^{}_* \cr
\sqrt{\frac{1}{6}} \hspace{0.1cm} c^{}_* - \sqrt{\frac{2}{3}} \hspace{0.1cm} s^{}_* e^{-i\phi} & \sqrt{\frac{1}{3}} \hspace{0.1cm} c^{}_* + \sqrt{\frac{1}{3}} \hspace{0.1cm} s^{}_* e^{-i\phi} &
\sqrt{\frac{1}{2}} \hspace{0.1cm} c^{}_* \cr
- \sqrt{\frac{1}{6}} & - \sqrt{\frac{1}{3}} & \sqrt{\frac{1}{2}} \cr }
\right ) P^{}_\nu \; ,
\end{eqnarray}
where $c^{}_*$ and $s^{}_*$ have been given below Eq. (11).
It is obvious that this ansatz can predict
$\theta^{}_l = \theta^{}_* \simeq 9.7^\circ$,
$\theta^{}_\nu = 45^\circ - \theta^{}_* \simeq 35.3^\circ$ and
$\theta = 45^\circ$. The Jarlskog invariant of leptonic CP violation
turns out to be
\begin{equation}
{\cal J} = c^{}_l s^{}_l c^{}_\nu s^{}_\nu c s^2 \sin\phi
= \frac{1}{36} \sin\phi \lesssim 0.028 \sin\phi
\end{equation}
in this scenario, and its maximal magnitude corresponds to
$\phi =\pm 90^\circ$.

Translating the results of three neutrino mixing angles
from the FX parametrization to the standard parametrization in
scenario B, we arrive at
\begin{eqnarray}
\theta^{}_{13} & = & \arcsin\left(\sqrt{\frac{1}{2}} \hspace{0.1cm}
s^{}_* \right) = \arcsin\left[\frac{1}{2 \left(\sqrt{3} + \sqrt{6} \right)}
\right] \simeq 6.9^\circ \; ,
\nonumber \\
\theta^{}_{23} & = & \arctan\left( c^{}_* \right) =
\arctan\left(\frac{1}{2\sqrt{3} - \sqrt{6}} \right) \simeq
44.6^\circ \; ,
\end{eqnarray}
and
\begin{eqnarray}
\theta^{}_{12} = \arctan\left[\sqrt{2} \left|\frac{1 - t^{}_* e^{i\phi}}
{2 + t^{}_* e^{i\phi}}\right|\right] \in \left[
\arctan\left(\frac{4}{6 + \sqrt{2}}\right) ,
\arctan\left(\frac{4}{3 + \sqrt{2}}\right) \right]
\simeq \left[28.3^\circ , 42.2^\circ \right]
\end{eqnarray}
with $t^{}_* \equiv \tan\theta^{}_* = (\sqrt{2} -1)/(\sqrt{2} +1)$. We
observe that the value of $\theta^{}_{12}$ in the standard parametrization
depends on the CP-violating phase $\phi$ in the FX parametrization,
and its minimal (or maximal) value corresponds to $\phi = 0^\circ$
(or $\phi = 180^\circ$). Once $\theta^{}_{12}$ is experimentally
determined to a good degree of accuracy, it will be possible to calculate
$\phi$ from Eq. (18). Given $\theta^{}_{12} \simeq 34.5^\circ$ \cite{GG}
for example,
\begin{equation}
\phi = \arccos\left[\frac{2 \left(1 + t^2_* \right) - \left(4 + t^2_*
\right) \tan^2\theta^{}_{12}}{4 \hspace{0.1cm} t^{}_* \left(1 + \tan^2\theta^{}_{12} \right)}\right] \simeq \pm 61.3^\circ \; ,
\end{equation}
which in turn leads to $|{\cal J}| \simeq 2.5\%$. This amount of CP
violation can in principle be measured in the future long-baseline
neutrino oscillation experiments. No doubt, the simplest experimental
way to distinguish between scenarios A and B is just to measure the
smallest neutrino mixing angle $\theta^{}_{13}$.

\vspace{0.5cm}

\framebox{\large\bf 4} ~
At present both scenarios A and B are compatible with the available
neutrino oscillation data. In either scenario the values of three
neutrino mixing angles are independent of four independent mass ratios
of charged leptons and neutrinos (i.e., $m^{}_e/m^{}_\mu$,
$m^{}_\mu/m^{}_\tau$, $m^{}_1/m^{}_2$ and $m^{}_2/m^{}_3$).
Of course, this kind of consequence is more or less contrived and it
implies some quite special textures of the charged-lepton and neutrino
mass matrices which might result from certain flavor symmetries \cite{Review}.
Instead of going into any details of model building, here we discuss a few
possibilities of reconstructing the charged-lepton mass matrix $M^{}_l$ and
the neutrino mass matrix $M^{}_\nu$ from a given pattern of the neutrino
mixing matrix $V$. For the sake of simplicity, we only concentrate on
scenario B to illustrate the salient features of our phenomenological treatment.

Let us define $M^{}_l$ and $M^{}_\nu$ in the following lepton mass terms by
assuming neutrinos to be the Majorana particles:
\begin{equation}
-{\cal L}^{}_{\rm mass} = \overline{\left(
\matrix{e & \mu & \tau}\right)^{}_{\rm L}} \hspace{0.1cm} M^{}_l
\left(\matrix{e \cr \mu \cr \tau}\right)^{}_{\rm R} +
\frac{1}{2} \hspace{0.1cm} \overline{\left(
\matrix{\nu^{}_e & \nu^{}_\mu & \nu^{}_\tau}\right)^{}_{\rm L}}
\hspace{0.1cm} M^{}_\nu \left(\matrix{\nu^c_e \cr \nu^c_\mu \cr \nu^c_\tau}\right)^{}_{\rm R} + {\rm h.c.} \; ,
\end{equation}
where $M^{}_\nu$ is symmetric but $M^{}_l$ is arbitrary.
One may diagonalize $M^{}_l$ and $M^{}_\nu$ via the transformations
$O^\dagger_l M^{}_l O^\prime_l = \widehat{M}^{}_l \equiv
{\rm Diag}\{m^{}_e, m^{}_\mu, m^{}_\tau\}$
and $O^\dagger_\nu M^{}_\nu O^*_\nu = \widehat{M}^{}_\nu \equiv
{\rm Diag}\{m^{}_1, m^{}_2, m^{}_3\}$,
where $O^{}_l$, $O^\prime_l$ and $O^{}_\nu$ are all unitary.
Then the flavor mixing matrix $V \equiv O^\dagger_l O^{}_\nu$ will
show up in the weak charged-current interactions
\begin{equation}
-{\cal L}^{}_{\rm cc} = \frac{g}{\sqrt{2}} \hspace{0.1cm} \overline{\left(
\matrix{e^\prime & \mu^\prime & \tau^\prime}\right)^{}_{\rm L}}
\hspace{0.1cm} \gamma^\mu V \left(\matrix{\nu^{}_1 \cr \nu^{}_2 \cr \nu^{}_3}\right)^{}_{\rm L} W^-_\mu + {\rm h.c.} \; ,
\end{equation}
where $\alpha^\prime$ (for $\alpha = e, \mu, \tau$) and $\nu^{}_i$
(for $i=1,2,3$) stand respectively for the mass eigenstates of charged
leptons and neutrinos. Now that both charged-lepton and neutrino
sectors may in general contribute to $V$, the reconstruction of $M^{}_l$
and $M^{}_\nu$ crucially depends on how one decomposes a given pattern
of $V$ into $O^{}_l$ and $O^{}_\nu$. Taking scenario B for example, we
consider four typical possibilities as follows.

{\it Possibility (1)}: $O^{}_l = {\bf 1}$ and $O^{}_\nu = V$. In this
case the mass eigenstates of three charged leptons are identified with
their flavor eigenstates, and hence only the neutrino sector is responsible
for the effect of flavor mixing. So we have $M^{}_l = \widehat{M}^{}_l$
and $M^{}_\nu = V \widehat{M}^{}_\nu V^T$. Most authors have taken such a flavor basis for building phenomenological models of $M^{}_\nu$ \cite{Review}.

{\it Possibility (2)}: $O^{}_\nu = {\bf 1}$ and $O^{}_l = V^\dagger$.
In this case the mass eigenstates of three neutrinos are identified with
their flavor eigenstates, and hence only the charged-lepton sector
contributes to flavor mixing.
So we have $M^{}_\nu = \widehat{M}^{}_\nu$
and $M^{}_l = V^\dagger \widehat{M}^{}_l {O^\prime_l}^\dagger$.
Because $O^\prime_l$ is in general unknown, it is actually difficult
to fix the texture of $M^{}_l$ in this flavor basis. One usually
assumes $M^{}_l$ to be Hermitian or symmetric so as to reduce the
number of degrees of freedom. If $M^{}_l$ is Hermitian, it can be
diagonalized via $O^\dagger_l M^{}_l O^{}_l = \widehat{M}^\prime_l \equiv
{\rm Diag}\{\lambda^{}_e, \lambda^{}_\mu, \lambda^{}_\tau\}$ with
$|\lambda^{}_\alpha| = m^{}_\alpha$ (for $\alpha =, e, \mu, \tau$).
Then it is possible to reconstruct $M^{}_l$ through
$M^{}_l = V^\dagger \widehat{M}^\prime_l V$. If $M^{}_l$ is symmetric,
we simply set $O^\prime_l = O^*_l = V^T$ and then arrive at
$M^{}_l = V^\dagger \widehat{M}^{}_l V^*$. A simple example of
this kind was originally given in Ref. \cite{FX96}.

{\it Possibility (3)}:
$O^{}_l = P^{}_\phi \otimes R^\dagger_{12}(\theta^{}_*)$
and $O^{}_\nu = R^{}_{23}(45^\circ) \otimes
R^{}_{12}(45^\circ -\theta^{}_*) \otimes P^{}_\nu$. Since $P^\dagger_\phi$
commutes with $R^{}_{23}(45^\circ)$, the product $O^\dagger_l O^{}_\nu$
automatically reproduces the pattern of $V$ shown in Eq. (15). Assuming
$M^{}_l$ to be symmetric for simplicity, we obtain
\begin{eqnarray}
M^{}_l & = & P^{}_\phi R^\dagger_{12}(\theta^{}_*) \widehat{M}^{}_l R^*_{12}(\theta^{}_*) P^T_\phi \; ,
\nonumber \\
M^{}_\nu & = & R^{}_{23}(45^\circ) R^{}_{12}(45^\circ -\theta^{}_*) P^{}_\nu \widehat{M}^{}_\nu P^T_\nu R^T_{12}(45^\circ -\theta^{}_*)
R^T_{23}(45^\circ) \; .
\end{eqnarray}
To be more explicit,
\begin{eqnarray}
M^{}_l & = & \left( \matrix{[m^{}_e c^2_* + m^{}_\mu s^2_* ]
e^{2i\phi} & [m^{}_e - m^{}_\mu] c^{}_* s^{}_* e^{i\phi} & 0
\cr \vspace{-0.4cm} \cr
[m^{}_e - m^{}_\mu] c^{}_* s^{}_* e^{i\phi} &
m^{}_e s^2_* + m^{}_\mu c^2_* & 0
\cr \vspace{-0.4cm} \cr
0 & 0 & m^{}_\tau} \right) \; ,
\nonumber \\
M^{}_\nu & = & \left( \matrix{ \frac{2}{3} \tilde{m}^{}_1
+ \frac{1}{3} \tilde{m}^{}_2 & \frac{1}{3} \left[ \tilde{m}^{}_2
- \tilde{m}^{}_1 \right] & \frac{1}{3} \left[ \tilde{m}^{}_1 -
\tilde{m}^{}_2 \right] \cr \vspace{-0.4cm} \cr
\frac{1}{3} \left[ \tilde{m}^{}_2
- \tilde{m}^{}_1 \right] & \frac{1}{6} \tilde{m}^{}_1 +
\frac{1}{3} \tilde{m}^{}_2 + \frac{1}{2} m^{}_3 & \frac{1}{2} m^{}_3
- \frac{1}{6} \tilde{m}^{}_1 - \frac{1}{3} \tilde{m}^{}_2 \cr
\vspace{-0.4cm} \cr
\frac{1}{3} \left[ \tilde{m}^{}_1 - \tilde{m}^{}_2 \right] &
\frac{1}{2} m^{}_3 - \frac{1}{6} \tilde{m}^{}_1 - \frac{1}{3}
\tilde{m}^{}_2 & \frac{1}{6} \tilde{m}^{}_1 + \frac{1}{3} \tilde{m}^{}_2
+ \frac{1}{2} m^{}_3 \cr} \right ) \; ,
\end{eqnarray}
where $\tilde{m}^{}_1 \equiv m^{}_1 e^{2i\rho}$ and
$\tilde{m}^{}_2 \equiv m^{}_2 e^{2i\sigma}$. It is interesting to notice
that the Dirac CP-violating phase $\phi$ is attributed to $M^{}_l$ while
the Majorana CP-violating phases $\rho$ and $\sigma$ come from $M^{}_\nu$
in this decomposition. One may derive both the textures of
$M^{}_l$ and $M^{}_\nu$ from certain flavor symmetries \cite{Review}.
For instance, the non-Abelian $A^{}_4$ flavor symmetry has been used to
derive the form of $M^{}_\nu$ in Eq. (23) \cite{AF}.

{\it Possibility (4)}:
$O^{}_l = P^{}_\phi \otimes R^\dagger_{23}(\theta^{}_*) \otimes
R^\dagger_{12}(\theta^{}_*)$ and
$O^{}_\nu = R^{}_{23}(45^\circ - \theta^{}_*) \otimes
R^{}_{12}(45^\circ - \theta^{}_*) \otimes P^{}_\nu$. Here again
the product $O^\dagger_l O^{}_\nu$ can reproduce the pattern of $V$
in Eq. (15). Assuming $M^{}_l$ to be symmetric, we analogously arrive at
\begin{eqnarray}
M^{}_l & = & P^{}_\phi R^\dagger_{23}(\theta^{}_*)
R^\dagger_{12}(\theta^{}_*) \widehat{M}^{}_l R^*_{12}(\theta^{}_*)
R^*_{23}(\theta^{}_*) P^T_\phi \; ,
\nonumber \\
M^{}_\nu & = & R^{}_{23}(45^\circ -\theta^{}_*)
R^{}_{12}(45^\circ -\theta^{}_*) P^{}_\nu \widehat{M}^{}_\nu P^T_\nu R^T_{12}(45^\circ -\theta^{}_*)
R^T_{23}(45^\circ -\theta^{}_*) \; .
\end{eqnarray}
More explicitly, we have
\begin{eqnarray}
M^{}_l & = & \left( \matrix{[m^{}_e c^2_* + m^{}_\mu s^2_* ]
e^{2i\phi} & [m^{}_e - m^{}_\mu] c^{2}_* s^{}_* e^{i\phi} &
[m^{}_e - m^{}_\mu] c^{}_* s^{2}_* e^{i\phi}
\cr \vspace{-0.4cm} \cr
[m^{}_e - m^{}_\mu] c^{2}_* s^{}_* e^{i\phi} &
[m^{}_e s^2_* + m^{}_\mu c^2_*] c^2_* + m^{}_\tau s^2_* &
[m^{}_e s^2_* + m^{}_\mu c^2_* - m^{}_\tau] c^{}_* s^{}_*
\cr \vspace{-0.4cm} \cr
[m^{}_e - m^{}_\mu] c^{}_* s^{2}_* e^{i\phi} &
[m^{}_e s^2_* + m^{}_\mu c^2_* - m^{}_\tau] c^{}_* s^{}_* &
[m^{}_e s^2_* + m^{}_\mu c^2_*] s^2_* + m^{}_\tau c^2_*} \right) \; ,
\nonumber \\
M^{}_\nu & = & \left( \matrix{ \frac{2}{3} \tilde{m}^{}_1
+ \frac{1}{3} \tilde{m}^{}_2 & \frac{2}{3} \sqrt{\frac{1}{3}} \left[ \tilde{m}^{}_2 - \tilde{m}^{}_1 \right] &
\frac{1}{3} \sqrt{\frac{2}{3}} \left[ \tilde{m}^{}_1 -
\tilde{m}^{}_2 \right] \cr \vspace{-0.4cm} \cr
\frac{2}{3} \sqrt{\frac{1}{3}} \left[ \tilde{m}^{}_2
- \tilde{m}^{}_1 \right] & \frac{2}{9} \tilde{m}^{}_1 +
\frac{4}{9} \tilde{m}^{}_2 + \frac{1}{3} m^{}_3 &
\frac{\sqrt{2}}{3} m^{}_3 - \frac{\sqrt{2}}{9} \tilde{m}^{}_1
- \frac{2\sqrt{2}}{9} \tilde{m}^{}_2 \cr
\vspace{-0.4cm} \cr
\frac{1}{3} \sqrt{\frac{2}{3}}
\left[ \tilde{m}^{}_1 - \tilde{m}^{}_2 \right] &
\frac{\sqrt{2}}{3} m^{}_3 - \frac{\sqrt{2}}{9} \tilde{m}^{}_1
- \frac{2\sqrt{2}}{9} \tilde{m}^{}_2 &
\frac{1}{9} \tilde{m}^{}_1 + \frac{2}{9} \tilde{m}^{}_2
+ \frac{2}{3} m^{}_3 \cr} \right ) \; ,
\end{eqnarray}
where $\tilde{m}^{}_1$ and $\tilde{m}^{}_2$ have been defined above.
In this case the textures of $M^{}_l$ and $M^{}_\nu$ are more or less
parallel to each other, implying that they could arise from a common
flavor symmetry or dynamic mechanism.

Indeed, there are infinite possibilities of reconstructing $M^{}_l$
and $M^{}_\nu$ for a given pattern of $V$ which is consistent with
current experimental data. From a phenomenological point of view,
we hope to make the textures of $M^{}_l$ and $M^{}_\nu$ as simple as
possible, or much easier to link with an underlying flavor symmetry.
In this sense the above examples just serve for illustration. Theoretically,
a viable neutrino mass model should predict or constrain the
proper forms of $M^{}_l$ and $M^{}_\nu$ from which the flavor mixing
matrix $V$ can be derived. But the inverse approach discussed above (i.e.,
starting from $V$ to reconstruct the textures of $M^{}_l$ and $M^{}_\nu$
based on a few assumptions) remains very useful because it is at least
possible to help give a ballpark estimate of the flavor structure that
a viable model ought to possess.

\vspace{0.5cm}

\framebox{\large\bf 5} ~
We have paid our attention to how to confront a constant neutrino mixing
pattern, which may be motivated by a certain flavor symmetry and can
predict $\theta^{}_{13} =0^\circ$ in the symmetry limit, with a
relatively large value of $\theta^{}_{13}$
(e.g., $5^\circ \lesssim \theta^{}_{13} \lesssim 12^\circ$). The latter
seems quite possible, at least not to be impossible, according to
some preliminary experimental hints extracted from current neutrino
oscillation data. We have shown that reasonable perturbations to the
democratic mixing pattern $U^{}_0$, which is geometrically related to 
the tri-bimaximal mixing pattern $V^{}_0$ through an
equal shift $\theta^{}_* \simeq 9.7^\circ$ of two large mixing
angles, can naturally produce a nearly tri-bimaximal neutrino
mixing matrix $V$ with relatively large $\theta^{}_{13}$.
We have proposed two simple but viable scenarios of $V$ for illustration:
one of them is based on the standard parametrization of $U^{}_0$ and
predicts $\theta^{}_{13} \simeq 9.7^\circ$, and the other relies on
the FX parametrization of $U^{}_0$ and predicts
$\theta^{}_{13} \simeq 6.9^\circ$. Both scenarios are in good
agreement with current neutrino oscillation data, and they can soon be
tested in a variety of more accurate neutrino oscillation experiments.

In this work we have tried not to go into any details of model building.
But we have discussed a few phenomenological possibilities of reconstructing
the charged-lepton mass matrix $M^{}_l$ and the neutrino mass matrix
$M^{}_\nu$ for a given neutrino mixing matrix $V$ with a relatively large
value of $\theta^{}_{13}$. Both the charged-lepton and neutrino sectors may
in general have significant contributions to $V$, and hence a specific
lepton flavor model should be able to determine the textures of $M^{}_l$
and $M^{}_\nu$ so as to give testable predictions for both neutrino mixing
angles and CP-violating phases.

Finally, we stress that it is not impossible to obtain a sufficiently
large value of $\theta^{}_{13}$ at the electroweak scale from finite
quantum corrections to a given constant neutrino mixing pattern with
$\theta^{}_{13} = 0^\circ$ \cite{Araki}. It is also possible to
generate $\theta^{}_{13}$ via the renormalization-group running
effects from the conventional seesaw 
scales of ${\cal O}(10^{14})$ GeV down to the electroweak scale
\cite{Antusch}, in particular when the seesaw threshold effects are 
taken into account \cite{Mei}.
But it seems more likely to achieve relatively large $\theta^{}_{13}$
from relatively significant symmetry breaking terms at a given scale
where the constant neutrino mixing pattern can be derived on the basis
of a certain flavor symmetry. In this sense our speculations and
discussions are expected to be phenomenologically useful and suggestive.

\vspace{0.5cm}

{\it Acknowledgments:}
I am indebted to K.K. Phua for his warm hospitality and
H. Fritzsch for sharing many joys of life at the IAS of NTU in Singapore,
where this paper was written. I am also grateful to T. Araki,
Y. Koide, and Y.F. Li for discussions.
This work was supported in part by the National Natural
Science Foundation of China under grant No. 10875131.

\begin{figure*}[t]
\centering \vspace{2cm}
\includegraphics[width=110mm]{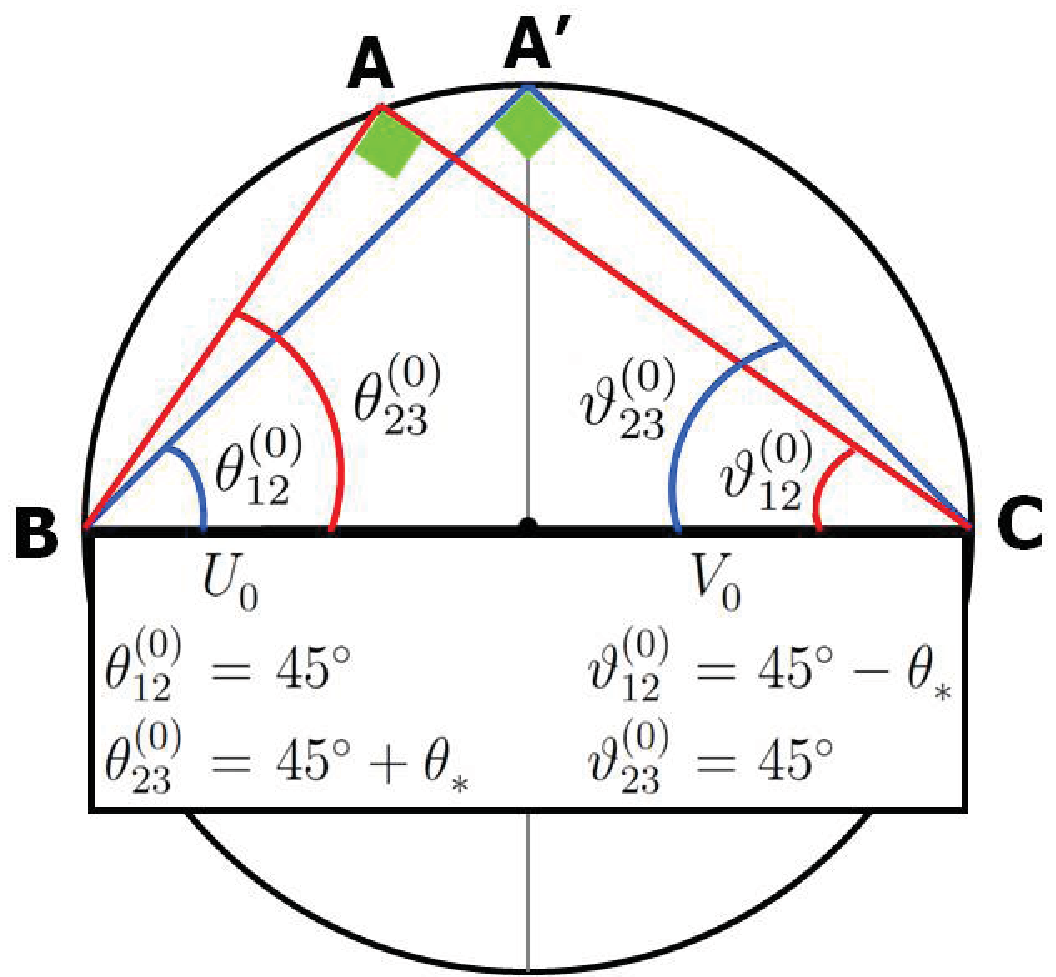}
\vspace{0.7cm} \caption{A geometric relationship between the
democratic mixing pattern $U^{}_0$ (with $\theta^{(0)}_{12} =
45^\circ$ and $\theta^{(0)}_{23} = 45^\circ + \theta^{}_*$) and the
tri-bimaximal mixing pattern $V^{}_0$ (with
$\vartheta^{(0)}_{12} = 45^\circ - \theta^{}_*$ and $\vartheta^{(0)}_{23} =
45^\circ$), where $\theta^{}_* = \arctan(\sqrt{2}) - 45^\circ = 45^\circ -
\arctan(1/\sqrt{2}) \simeq 9.7^\circ$. Four nonzero mixing angles
of $U^{}_0$ and $V^{}_0$ correspond to four inner angles of two
right triangles $\triangle {\rm ABC}$ and $\triangle {\rm A^\prime
BC}$ in the real plane.}
\end{figure*}

\end{document}